\documentclass[11pt]{article}
\usepackage[dvips]{graphics}

\title{There is no first quantization -- except in the 
de Broglie--Bohm interpretation}

\author{Hrvoje Nikoli\'c \\
Theoretical Physics Division, Rudjer Bo\v{s}kovi\'{c} Institute, \\
P.O.B. 180, HR-10002 Zagreb, Croatia \\
{\normalsize hrvoje@thphys.irb.hr} \\
\makebox[1in]{} \\
}
\date{\today}
\begin{document}
\maketitle
\begin{abstract}
The relativistic effects of the integer-spin quantum field theory 
imply that the wave functions describing a fixed number of 
particles do not admit the usual probabilistic interpretation. 
Among several most popular interpretations of quantum mechanics
applied to first quantization, 
the only interpretation for which this fact does not lead to 
a serious problem,  
and therefore the only consistent interpretation
of first quantization, is the de Broglie--Bohm interpretation. 
\end{abstract}
\vspace*{0.5cm}
\vspace*{0.5cm}

\noindent
Recently, 
Zeh argued that ``there is no first quantization" \cite{zeh}, 
i.e. that true quantum theory is ``second" 
quantization, known also under the name {\em quantum field theory} 
(QFT). According to this paradigm, particles appear as 
small localized objects 
not because the fundamental objects are quantized pointlike 
objects called particles, but rather due to decoherence caused 
by local interactions between fields.

In this Letter we present further arguments 
against first quantization, by emphasizing
that the usual postulates and interpretations of quantum mechanics 
(QM) applied to first quantization are inconsistent with 
relativistic QFT.
Specifically, we consider the Copenhagen interpretation,
the minimal probabilistic interpretation, 
the consistent-history interpretation,
the Everettinterpretation and
the de Broglie--Bohm interpretation, and emphasize that all these 
interpretations except the last one require one of the basic 
postulates of QM: the
existence of a quantity that can be 
naturally interpreted as a probability. 
On the other hand, from relativistic QFT it follows that this 
postulate cannot be applied to wave functions that describe 
a fixed number of particles. Therefore, as far as {\em only  
first} quantization is considered, only 
two points of view seem to be consistent. One consistent 
point of view is that 
decoherence is the best we can say about the problem of 
measurement. (The decoherence paradigm \cite{zeh2,joos,zur,omn}
is not an interpretation at all;
decoherence is a physical phenomenon 
for which there are no doubts that it
occurs in nature.) The other consistent point of view is the 
de Broglie--Bohm interpretation \cite{bohm,bohmPR1,bohmPR2,
holPR,holbook} 
formulated in a form in 
which both the fields and the particle trajectories are 
physical entities that objectively exist \cite{nikol1,nikol2}.        
   
Before discussing various interpretations of QM, let us first 
state a few basic phenomenological rules of QM that are 
experimentally confirmed
and thus should be consistent with any consistent interpretation, 
at least in the cases for which the experimental confirmation 
exists. Let $X$ denote some set of the degrees of freedom, say 
positions of $n$ particles. Assume 
that a wave function $\psi(X,t)$ can be rewritten in terms of 
some functions $\psi_i(X,t)$ as a linear combination
\begin{equation}\label{eq1}
\psi(X,t)=\sum_i c_i(t) \psi_i(X,t),
\end{equation}
where $c_i(t)$ are complex coefficients. Assume also that a 
unitary scalar product $(\; , \;)$ of wave 
functions can be defined such that
\begin{equation}\label{eq2}
(\psi_i,\psi_j)=\delta_{ij},
\end{equation}
independently of time $t$.
Assume further that there 
exists a hermitian operator $\hat{A}$ acting on the $X$-space, such that
\begin{equation}\label{eq3}
\hat{A}\psi_i(X,t)=a_i(t)\psi_i(X,t),
\end{equation}
where $a_i(t)$ are real functions. {\em If} all these assumptions
are fulfilled, then we can apply the following rules:
\begin{enumerate}
\item $\hat{A}$ represents an observable and
the probability of finding the system at time $t$ 
to have the value $a_i(t)$ 
of the observable $\hat{A}$ is equal to $|c_i(t)|^2$.
\item If the sistem is found to have the value $a_i$ of the 
observable $\hat{A}$ at time $t$, then the probabilities at 
later times should be calculated by replacing $\psi$ with 
$\psi_i$. 
\end{enumerate}
Note that, by stating the rules above, we do not claim that 
the mentioned assumptions {\em are} fulfilled. 
Instead, we merely 
state what we may conclude {\em if} they are fulfilled. These 
rules say nothing about the case in which the mentioned 
assumptions are not fulfilled.

Let us now shortly discuss particular interpretations. (A more detailed
review of all most popular interpretations of QM can be found in 
\cite{lal}.) All these interpretations are attempts to formulate 
{\em fundamental principles valid in all physically possible cases} 
from which the phenomenological rules above can be derived. 

We start with the Copenhagen interpretation. The claim that 
$|c_i(t)|^2$ represent probabilities is one of its fundamental 
postulates. This postulate implies that a unitary scalar product with the 
property (\ref{eq2}) {\em must} exist, 
which is related to the fact 
that probabilities cannot be negative and that the sum 
$\sum_i |c_i(t)|^2$ of all
probabilites, at each time, must be equal to 1. 
Another fundamental postulate says 
that, in the process of measurement, the wave function $\psi$ 
``collapses" into $\psi_i$, i.e. after the measurement 
the wave function becomes {\em equal} to $\psi_i$. 

The minimal probabilistic interpretation avoids the introduction 
of a notorious wave-function-collapse postulate \cite{lal}. However, 
similarly to the Copenhagen interpretation, it postulates that
$|c_i(t)|^2$ represent probabilities, which implies that a 
unitary scalar product with the property (\ref{eq2}) must exist.           

The basic objects in the consistent-history interpretation 
\cite{grif,omn2,gell} are
sequences of projection operators at different times --
called histories. The 
consistent-history interpretation generalizes the usual 
rules for calculating probabilities in QM by postulating 
the rules for calculating the probabilities of
histories. 
Therefore, the consistent-history interpretation also 
relies on the assumption of the 
consistency of the probabilistic interpretation.

The Everett relative-state interpretation \cite{ever,wheel}
(some variants 
of which are also known under the name ``many-worlds interpretation"
\cite{ben}) avoids a postulation of an {\it a priori} probabilistic 
interpretation. The only fundamental postulate in this interpretation 
is the usual causal evolution of the wave function. 
Therefore, this is a completely deterministic theory. 
However, this is not a 
deterministic theory of observables. Instead, this is merely a 
deterministic theory of wave functions.
Therefore, in this interpretation, 
the explanation of the rule that $|c_i(t)|^2$ represent 
the probabilities 
is a very controversial subject \cite{ben}. In order to explain it,  
some researchers use rather indirect and vague arguments \cite{ever,teg}
or base their arguments on additional assumptions \cite{deut,wall,saun}, 
while others conclude (in our opinion, correctly) 
that the probabilistic rule cannot be explained 
in the framework of the Everett interpretation \cite{bal,kent,bar}.       
What is even more important for our discussion, 
all existing arguments for the probabilistic rule 
are based on the assumptions that we mentioned before expressing 
this rule.
In the framework of the Everett interpretation,
it is not clear at all how one might, even in principle, 
calculate the probabilities when these assumptions are not 
fulfilled.

The de Broglie--Bohm (dBB) interpretation 
\cite{bohm,bohmPR1,bohmPR2,holPR,holbook} 
is a deterministic nonlocal hidden-variable interpretation.
In contrast with the Everett interpretation, it is not only 
a deterministic theory of wave functions, but also a deterministic 
theory of observables.
In the case of nonrelativistic first quantization, 
its basic postulate is that the particle velocities are given 
by (we take $\hbar=1$)
\begin{equation}\label{dbb}
V=\frac{1}{m}\partial_X S(X,t),
\end{equation}
where $m$ is the mass of a particle and $S$ is the phase of the 
wave function $\psi(X,t)=R(X,t)e^{iS(X,t)}$. No additional 
assumption is required for the properties of $\psi(X,t)$. 
Instead, it is rather a {\em prediction} of this theory that 
{\em if} $\psi$ satisfies the usual nonrelativistic 
Schr\"odinger equation of the form
\begin{equation}\label{5'}
-\frac{\nabla^2\psi}{2m}+U(X)\psi=-i\partial_t\psi,
\end{equation}
which provides that
\begin{equation}\label{cons}
\frac{d}{dt}\int d^{3n}\!X\, |\psi(X,t)|^2 =0,
\end{equation}
{\em then} the statistical distribution of particles $\rho$ 
is given by
\begin{equation}\label{distrib}
\rho(X,t)=|\psi(X,t)|^2
\end{equation}
for every $t$, 
provided that at an initial time $t_0$
\begin{equation}\label{2post}
\rho(X,t_0)=|\psi(X,t_0)|^2.
\end{equation} 
Eq.~(\ref{2post}) is also often postulated as one of the fundamental 
postulates of the dBB interpretation, but this postulate is not 
really necessary. Instead, one can show that if
(\ref{5'}) is satisfied and the interactions are sufficiently 
complex, then the distribution (\ref{distrib}) is the most probable
\cite{bohm2,val} (see also \cite{durr,durr'}). 

It is, of course, well known that the probability distribution
(\ref{distrib}) is a special case of the general quantum mechanical 
rule for calculating probabilities explained above. This is most 
easily seen by writing (\ref{eq1}) as
\begin{equation}
\psi(X,t)=\int d^{3n}\!X'\, \psi(X',t) \delta^{3n}(X'-X).
\end{equation}
The scalar product is
\begin{equation}
(\psi_i,\psi_j)=\int d^{3n}\!X\,\psi_i^*(X,t)\psi_j(X,t),
\end{equation}
which is time independent, owing to (\ref{5'}).
Since the quantity $(\psi_i,\psi_i)$ is positive definite, 
the functions $\psi_i$ can be chosen such that 
(\ref{eq2}) is satisfied. 
Therefore, we see that the dBB interpretation {\em explains}, 
rather than postulates, the rules for calculating probabilities 
that particles are at certain positions. 

Now consider {\em relativistic} QM. In a relativistic generalization 
of the Schr\"odinger equation, the wave function of 
a single free particle with an integer spin
satisfies the Klein-Gordon equation (we take $\hbar=c=1$)
\begin{equation}\label{KG}
(\partial_t^2-\nabla^2+m^2)\psi({\bf x},t)=0.
\end{equation}
It is well known that, owing to the second time derivative in (\ref{KG}), 
$\psi$ does not satisfy (\ref{cons}). Therefore, the quantity 
$|\psi|^2$ cannot be interpreted as a probability density. 
If, in (\ref{cons}), $|\psi|^2$ is replaced by 
$j_0\equiv i(\psi^*\partial_t \psi - \psi\partial_t\psi^*)$, then such a 
modified Eq.~(\ref{cons}) is satisfied. However, the density 
$j_0$ is not positive definite, so it cannot be interpreted as a 
probability density either. The time-independent scalar product is
\begin{equation}
(\psi_i,\psi_j)=i\int d^3x(\psi_i^*\partial_t\psi_j
-\psi_j\partial_t\psi_i^*) .
\end{equation}
The quantity $(\psi_i,\psi_i)$ is not positive definite, 
which is in contradiction with the assumption (\ref{eq2}).
A consistent interpretation of $\psi$ is 
given by second quantization, i.e. QFT (see e.g. \cite{bjor}).   
In QFT, (\ref{KG}) becomes an equation for the operator $\hat{\psi}$.
The quantity $\psi$ is not a wave function but an observable 
attributed to the operator $\hat{\psi}$. The wave functional 
$\Psi[\psi({\bf x}),t]$ determines probabilities of these observables 
according to the usual quantum mechanical rule. An $n$-particle 
wave function can be defined as a matrix element of the form 
$\langle 0|\hat{\psi}({\bf x}_1,t)\cdots\hat{\psi}({\bf x}_n,t)|\Psi
\rangle$. From Eq.~(\ref{KG}) applied to $\hat{\psi}$ one can see 
that wave functions satisfy multiparticle generalizations of 
(\ref{KG}), which are
equations with the second time derivatives. 

Now we see that interpretations of first quantization that 
contain the probabilistic 
interpretation as one of its fundamental postulates are not consistent. 
One can propose that such interpretations are valid only 
in the case when the relativistic effects 
are negligible, but in that case, such interpretations cannot be 
viewed as fundamental principles. Instead, they should be 
consequences of some more fundamental principles. If one postulates 
that the probabilistic interpretation is fundamental only at the level of 
wave functionals of QFT, 
then, even in the nonrelativistic limit,
it does {\em not} follow at all from this interpretation
that the probabilistic interpretation can be applied at the level 
of $n$-particle wave functions.     

There are 3 possible ways to cope with this problem. 
The first approach is to 
take the rules of first quantization as 
unfundamental phenomenological rules 
valid only under restricted conditions, without wondering about 
the true reasons why these rules are valid. 
The second approach, 
probably the most popular one at the moment, is to view these rules as 
consequences of the decoherence at the level of true quantum 
degrees of freedom -- fields \cite{zeh}. 
This approach is so popular because it explains, rather than postulates, 
how the quantum interaction with a sufficiently complex environment 
forces the state of a system to ``collapse" to a state that 
corresponds to a diagonal (i.e. classical) density matrix, 
using only the properties of solutions of the Schr\"odinger equation or 
any of its generalizations. 
However, as admitted by most experts for decoherence, it 
does not explain the ``collapse" of the wave function. In 
other words, decoherence 
explains why we can use classical statistics
to statistically describe the outcome of quantum measurements, but 
it does not explain why and how the measured
observables take definite values.

The third possible approach is the dBB interpretation. 
Actually, it is important to emphasize that the usual form 
of the dBB interpretation \cite{bohm,bohmPR1,bohmPR2,holPR,holbook} 
applies the dBB interpretation either to particles or to fields, 
not to both. In fact, it is often interpreted as a theory 
in which, at a fundamental level, bosons are fields and fermions 
are particles. However, in a recent variant of the dBB interpretation 
both the fields and the particle trajectories simultaneously exist 
as objective entities \cite{nikol1,nikol2}. (A similar variant of 
the dBB interpretation is presented also in \cite{durr2}, but 
with an important difference that, contrary to 
the variant in \cite{nikol1,nikol2}, 
it is not relativistic and completely deterministic.) 
Therefore, the dBB interpretation 
can be formulated such that the usual rules for calculating 
probabilities in nonrelativistic first quantization can be 
{\em derived} from more fundamental principles that respect 
the relativistic effects. For a known wave function,
in general, it is not so easy to calculate the probabilities
of particle positions, except in the nonrelativistic limit. 
However, these probabilites can be calculated in principle 
from the deterministic 
theory of particle trajectories determined by a relativistic 
generalization of (\ref{dbb}).
In this sense, the relativistic 
generalization of the dBB interpretation of QM is able to give 
testable predictions on certain relativistic quantum 
phenomena on which the conventional theory 
does not give any prediction at all.    

We also want to emphasize that two consistent points of 
view: the decoherence paradigm and the dBB interpretation -- are not 
two competitive points of view. Instead, the decoherence is an 
important ingredient of the dBB interpretation, 
without which the dBB interpretation cannot explain the usual 
rules for calculating probabilities. 
To see why, note 
first that the dBB interpretation predicts the distribution of momenta 
to be given by 
\begin{equation}
\rho(P,t)=\int d^{3n}\!X\, |\psi(X,t)|^2 
\delta^{3n}(P-\partial_X S(X,t)),
\end{equation}
which, in general, is {\em not} equal to the quantum mechanical 
distribution $|\tilde{\psi}(P,t)|^2$ [where $\tilde{\psi}(P,t)$  
is the Fourier 
transform of $\psi(X,t)$]. However, this problem resolves 
when the general theory of
quantum measurements in the dBB interpretation 
\cite{bohm,bohmPR1,holbook} is taken into account. 
This theory of measurements has much in common 
with the theory of quantum measurements
based on decoherence, which is perhaps the clearest from the 
presentation in \cite{holbook}. In a way, the dBB interpretation 
can be viewed as a completion 
of the theory of measurements based on decoherence, because 
the dBB interpretation uses the basic ideas of decoherence, but 
at the same time provides an answer to the question that decoherence 
by itself cannot answer, namely why and how observables take definite 
values.

By taking into account the relativistic effects in free QFT,
in this Letter we have discredited all discussed interpretations 
of first quantization, except the dBB interpretation. 
Although it is not the aim of this Letter to discuss in detail 
the effects of field interactions, we 
note that the {\em linearity} of the Schr\"odinger equation 
is also one of the fundamental postulates in some interpretations 
of QM. On the other hand, the wave functions that result from 
{\em interacting} QFT do not satisfy linear equations. 
These nonlinearities do not lead to any fundamental problem in the 
dBB interpretation of first quantization \cite{nikol1,nikol2}.

To conclude, we repeat that among several most popular interpretations 
of QM we discussed, 
only the dBB interpretation can be applied to first quantization, 
because other interpretations rely on assumptions that are not really 
satisfied. Therefore, if any of these interpretations of first 
quantization is consistent, it can only be the dBB interpretation.
Of course, it is possible that some other interpretations 
of first quantization, not considered in this paper, are also 
consistent. In particular, it is certainly possible that some 
other deterministic hidden-variable interpretations, more complicated 
than the dBB interpretation, are consistent. However,
it was not possible to discuss all existing interpretations, 
so the discussion of this Letter is
restricted to the most popular ones. 

This work was supported by the Ministry of Science and Technology of the
Republic of Croatia under Contract No. 0098002.


\begin{thebibliography}{99}

\bibitem{zeh}
H.~D.~Zeh, Phys.~Lett.~A 309 (2003) 329.
\bibitem{zeh2} 
H.~D.~Zeh, Found.~Phys.~1 (1970) 69.
\bibitem{joos}
E.~Joos, H.~D.~Zeh, Z.~Phys.~B 59 (1985) 223.
\bibitem{zur}
W.~H.~Zurek, Phys.~Today 44 (1991) 36.
\bibitem{omn}
R.~Omn\`es, Phys.~Rev.~A 56 (1997) 3383.
\bibitem{bohm}
D.~Bohm, Phys.~Rev.~85 (1952) 166, 180.
\bibitem{bohmPR1}
D.~Bohm, B.~J.~Hiley, P.~N.~Kaloyerou,
Phys. Rep.~144 (1987) 323.
\bibitem{bohmPR2}
D.~Bohm, B.~J.~Hiley,
Phys.~Rep.~144 (1987) 349.
\bibitem{holPR}
P.~R.~Holland, Phys.~Rep.~224 (1993) 95.
\bibitem{holbook}
P.~R.~Holland, The Quantum Theory of Motion,
Cambridge University Press, Cambridge, 1993.
\bibitem{nikol1}
H.~Nikoli\'c, quant-ph/0208185.
\bibitem{nikol2}  
H.~Nikoli\'c, quant-ph/0302152.
\bibitem{lal}
F.~Lalo\"e, Am.~J.~Phys.~69 (2001) 655.
\bibitem{grif}
R.~B.~Griffiths, J.~Stat.~Phys.~36 (1984) 219.
\bibitem{omn2}
R.~Omn\`es, J.~Stat.~Phys.~53 (1988) 893.
\bibitem{gell}
M.~Gell-Mann, J.~Hartle, Phys.~Rev.~D 47 (1993) 3345.
\bibitem{ever}
H.~Everet, Rev.~Mod.~Phys.~29 (1957) 454.
\bibitem{wheel}
J.~A.~Wheeler, Rev.~Mod.~Phys.~29 (1957) 463.
\bibitem{ben}
Y.~Ben-Dov, Am.~J.~Phys.~58 (1990) 829.
\bibitem{teg}
M.~Tegmark, Fortsch.~Phys.~46 (1998) 855.
\bibitem{deut}
D.~Deutsch, Proc.~Roy.~Soc.~Lond.~A 455 (1999) 3129.
\bibitem{wall}
D.~Wallace, quant-ph/0211104; quant-ph/0303050.
\bibitem{saun}
S.~Saunders, quant-ph/0211138.
\bibitem{bal}
L.~E.~Ballentine, Found.~Phys.~3 (1973) 229.
\bibitem{kent}
A.~Kent, Int.~J.~Mod.~Phys.~A 5 (1990) 1745; gr-qc/9703089.
\bibitem{bar}
H.~Barnum, C.~Caves, J.~Finkelstein, C.~Fuchs, R.~Schack, 
Proc.~Roy.~Soc.~Lond.~A 456 (2000) 1175.
\bibitem{bohm2}
D.~Bohm, Phys.~Rev.~89 (1953) 458.
\bibitem{val}
A.~Valentini, Phys.~Lett.~A 156 (1991) 5.
\bibitem{durr}
D.~D\"urr, S.~Goldstein, N.~Zangh\'i,
J.~Stat.~Phys.~67 (1992) 843.
\bibitem{durr'}                     
D.~D\"urr, S.~Goldstein, N.~Zangh\'i,
Phys.~Lett.~A 172 (1992) 6.
\bibitem{bjor}
J.~D.~Bjorken, S.~D.~Drell, Relativistic Quantum Fields,
McGraw-Hill, New York, 1965.
\bibitem{durr2}
D.~D\"urr, S.~Goldstein, R. Tumulka, N.~Zangh\'i,
J.~Phys.~A 36 (2003) 4143; quant-ph/0303156.


\end{thebibliography}
\end{document}